\documentclass[prl,twocolumn,showpacs,preprintnumbers,amsmath,amssymb]{revtex4}
\usepackage{graphicx}
\usepackage{dcolumn}
\usepackage{bm}
\usepackage{mathrsfs}
\usepackage{subfigure}
\usepackage{natbib}
\usepackage[dvips]{color}
\usepackage[normalem]{ulem}

\begin{document}
\title{Two-band superconductors: Hidden criticality deep in the superconducting state}

\author{L. Komendov\'{a}}
\affiliation{Departement Fysica, Universiteit Antwerpen, Groenenborgerlaan 171, B-2020 Antwerpen, Belgium}

\author{Yajiang Chen}
\affiliation{Departement Fysica, Universiteit Antwerpen, Groenenborgerlaan 171, B-2020 Antwerpen, Belgium}

\author{A. A. Shanenko}
\affiliation{Departement Fysica, Universiteit Antwerpen, Groenenborgerlaan 171, B-2020 Antwerpen, Belgium}

\author{M. V. Milo\v{s}evi\'{c}}
\affiliation{Departement Fysica, Universiteit Antwerpen, Groenenborgerlaan 171, B-2020 Antwerpen, Belgium}

\author{F. M. Peeters}
\affiliation{Departement Fysica, Universiteit Antwerpen, Groenenborgerlaan 171, B-2020 Antwerpen, Belgium}
\date{\today}

\begin{abstract}
We show that two-band superconductors harbor hidden criticality deep in the superconducting state, stemming from the critical temperature of the weaker band taken as an independent system. For sufficiently small interband coupling $\gamma$ the coherence length of the weaker band exhibits a remarkable deviation from the conventional monotonic increase with temperature, namely, a pronounced peak
close to the hidden critical point. The magnitude of the peak scales as $\propto \gamma^{-\mu}$, with the Landau critical exponent $\mu =\frac{1}{3}$, the same as found for the mean-field critical behavior with respect to the source field in ferromagnets and ferroelectrics. Here reported hidden criticality of multi-band
superconductors can be experimentally observed by, e.g., imaging of the variations of the vortex core in a broader temperature range. Similar effects are expected for the superconducting multilayers.
\end{abstract}


\maketitle

Critical phenomena~\cite{lan} constitute one of the most important aspects of the physics of complex systems. The classical scenario of both thermal and quantum criticality involves the ordered phase induced by a breakdown of a basic symmetry and the disordered phase appearing due to the restoration of this symmetry. These two phases are separated by a critical point, i.e., a second-order phase transition accompanied by critical phenomena, e.g., the Curie point, the  superconducting-to-normal state transition, the metal-insulator transition etc. (see, e.g., \cite{lan,pok}). However, possible realizations of the critical behavior are not restricted to this standard picture. The present Letter reports a fascinating example when in addition to the standard critical behavior, the system is affected by {\it hidden criticality deep in the ordered phase}. This is the case of a two-band superconductor \cite{suhl} (for recent activity see, e.g., \cite{GL,mosh,zhit,kogan,shan,new}) where the interband coupling measures the proximity to the hidden critical point and plays the role of the source field governing the hidden criticality.

The family of multi-band superconducting materials is characterized by the presence of multiple sheets of the Fermi surface (bands). In this case the superconducting properties are controlled by a set of different band condensates \cite{suhl}. Due to the presence of a nonzero interband coupling, such condensates are not independent. As a result, a band order parameter, i.e., the measure of the
condensate in a given band, is not simply proportional to the Cooper-pair amplitude in this band but involves a sum over all band pairing amplitudes, each multiplied by a specific coefficient, i.e., the coupling constant $g_{ij}=g_{ji}$. In a two-band superconductor, when pairing of electrons between bands is negligible (e.g., due to symmetry reasons), the order parameter $\Delta_i({\bf x})$ reads
\begin{equation}
\Delta_i({\bf x})=\sum\limits_{j=1,2}\;g_{ij}\langle {\hat\psi}_{j
\uparrow}({\bf x}){\hat \psi}_{j\downarrow}({\bf x})\rangle,
\label{gap}
\end{equation}
where $i,j$ enumerate the two bands, and $\langle {\hat\psi}_{j\uparrow}({\bf x}){\hat \psi}_{j\downarrow}({\bf x})\rangle$ is the band-dependent Cooper pair amplitude (anomalous average of the field operators with the spin projection up and down for the singlet s-wave pairing). The system described by Eq.~(\ref{gap}) has a unique critical temperature $T_c$ for any nonzero interband coupling $g_{12}$. However, when $g_{12}=0$, one deals theoretically with two independent superconducting condensates with two different critical temperatures $T_{c1}$~(for the stronger band) and $T_{c2}$~(for the weaker band), and the basic symmetry changes from ${\mathrm U}(1)$ to ${\mathrm U}(1)\times {\mathrm U}(1)$. Here the question arises whether or not the behavior of the more realistic, \textit{weakly} coupled system is affected by the proximity of decoupled bands.

{\itshape Analytical results.} The initial step in our study is to get analytical information about a two-gap system with weakly coupled components. With this in mind, we examine Eq.~(\ref{gap}) at $T=T_{c2}$ by developing a Ginzburg-Landau (GL) type of approach for $g_{12} \to 0$. Actually, instead of $g_{12}$ it is more convenient to deal with the dimensionless parameter $\gamma =\lambda_{12}
/(\lambda_{11}\lambda_{22})$, where $\lambda_{ij}=g_{ij}N(0)$, and $N(0)$ is the total density of states at the Fermi energy, i.e., $N(0)=\sum_i N_i(0)$. The GL approach invokes an expansion in powers of the order parameter and its spatial derivatives. Such an expansion for the Cooper-pair amplitude (anomalous Green's function) in a two-band superconductor reads~\cite{zhit,kogan,shan}
\begin{eqnarray}
\frac{\langle {\hat\psi}_{i\uparrow}({\bf x}){\hat
\psi}_{i\downarrow}({\bf x})\rangle}{N(0)}=\alpha_i \Delta_i-\beta_i
\Delta_i|\Delta_i|^2+{\cal K}_i{\bf D}^2\Delta_i+\ldots,
\label{exp_anom}
\end{eqnarray}
with the gauge-invariant gradient ${\bf D}={\boldsymbol\nabla}-\frac{2ie}{\hbar c} {\bf A}$ and the parameters (in the clean limit)
\begin{eqnarray}
\alpha_i= n_i\ln\Big(\frac{2e^{\Gamma}\hbar\omega_c}{\pi
T}\Big),\;\beta_i =n_i\frac{7\zeta(3)}{8\pi^2T^2},\;{\cal
K}_i=\frac{\beta_i}{6}\hbar^2 v^2_i. \label{coeff}
\end{eqnarray}
Here $\hbar\omega_c$ is the cut-off energy, $\zeta(x)$ is the Riemann zeta-function, $\Gamma = 0.577$ is the Euler constant, $n_i=N_i(0)/N(0)$, and $v_i$ is the band-dependent Fermi velocity. Note however that at $T=T_{c2}$ the expansion given by Eq.~(\ref{exp_anom}) holds only for the weaker band in the limit $\gamma \to 0$, while it is inappropriate for the stronger band where the order parameter does not vanish in this limit. Nevertheless, Eq.~(\ref{exp_anom}) provides plenty of information about the weaker band. Using Eq.~(\ref{gap}) and invoking the expansion of Eq.~(\ref{exp_anom}) for the weaker band, we find~\cite{suppmat} the following $\gamma$GL equation for $\Delta_2$ at $T=T_{c2}$:
\begin{eqnarray}
\beta_2\Delta_2|\Delta_2|^2 -{\cal K}_2{\bf D}^2\Delta_2 -\gamma
\Delta_{1,\gamma\to0}= 0, \label{gGL}
\end{eqnarray}
with $\Delta_{1,\gamma \to 0}$ being the order parameter of the strong band in the limit $\gamma \to 0$, and $\beta_2$ and ${\cal K}_2$ given by Eq.~(\ref{coeff}) at $T=T_{c2}$. We stress that Eq.~(\ref{gGL}) is exact for $\Delta_2$ in the leading order in $\gamma$ at $T=T_{c2}$. Note that due to $T=T_{c2}$ there is no
linear term in $\Delta_2$ in Eq.~(\ref{gGL}), in contrast to the ordinary GL theory. Despite the absence of the linear term in $\Delta_2$, a stable solution to Eq.~(\ref{gGL}) exists due to the presence of the source term $\gamma \Delta_{1, \gamma \to 0}$, which reflects the Josephson-like coupling between the bands. This solution can be obtained only after solving the proper formalism for the stronger band at $\gamma = 0$. Nevertheless, qualitative information about $\Delta_2$ as a function of $\gamma$ can be found from a general analysis of Eq.~(\ref{gGL}). In the simplest case of a spatially uniform system we obtain for $\gamma\to 0$ and $T=T_{c2}$
\begin{eqnarray}
\Delta_2 = \Big(\frac{\gamma\Delta_{1,\gamma\to 0}}{\beta_2}\Big)^{1/3},
\;\frac{\partial\Delta_2}{\partial\gamma} = \frac{\gamma^{-2/3}}{3}
\Big(\frac{\Delta_{1,\gamma\to 0}}{\beta_2}\Big)^{1/3}.
\label{unif}
\end{eqnarray}
Based on Eq.~(\ref{unif}) and assuming a similar general dependence of $\Delta_2$ on $\gamma$, we can analyze the solution of Eq.~(\ref{gGL}) in the presence of a spatially nonuniform condensate. We find for $\gamma \to 0$ and $T=T_{c2}$
\begin{eqnarray}
\Delta_2 \propto \gamma^{1/3}, \;\; \frac{\partial \Delta_2}{\partial
\gamma} \propto \gamma^{-2/3},\;\; \xi_2 \propto \gamma^{-1/3},
\label{crit_beh}
\end{eqnarray}
where $\xi_i$ denotes the band-dependent healing (coherence) length and the asymptotic behavior for $\xi_2$ is found from $\nabla^2 \Delta_2\propto \xi_2^{-2} \Delta_2 \propto \gamma$. Equations~(\ref{unif}) and (\ref{crit_beh}) therefore reveal the critical behavior (see, e.g., Ref.~\onlinecite{lan}) of the order
parameter in the weaker band $\propto \gamma^{1/\delta}$, the corresponding susceptibility $\propto \gamma^{1/\delta-1}$ and the healing length (proportional to the correlation radius of fluctuations $r_c$) $\propto \gamma^{-\mu}$, with the Landau mean-field critical exponents $\delta = 3$ and $\mu=\frac{1}{3}$. As seen, the interband coupling $\gamma$ can be interpreted as a source field governing the hidden criticality in two-band superconductors~\cite{note_source}. This is analogous to the source fields in criticality of e.g. ferromagnets (magnetic field) or ferroelectrics (electric field), where the same Landau exponents are found within
the mean field. Here we note that $\lambda_{12}$ is not easily tunable in two-band superconductors,  which is different as compared to the source field in ferromagnets and ferroelectrics. However, the coexistence of two weakly coupled order parameters was also achieved in layered mesoscopic rings~\cite{bluhm}, where the interlayer coupling can be varied by changing the distance between two
participating layers. The physics of the coherent phenomena in such a two-layer system is essentially the same as in two-band superconductors~\cite{bluhm,decol}.

\begin{figure*}
\includegraphics[width=0.94\linewidth]{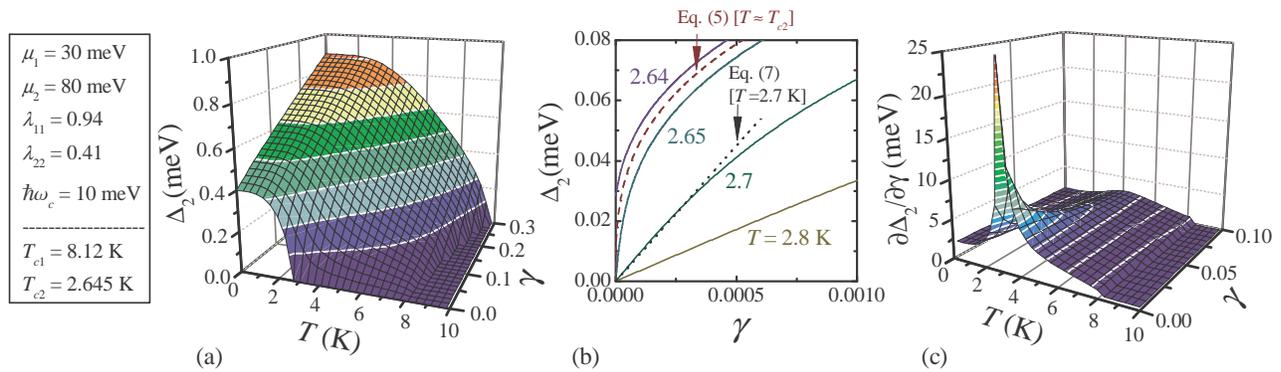}
\caption{(Color online) (a) The weaker-band order parameter $\Delta_2$ versus the temperature $T$ and the interband coupling $\gamma$ as calculated from the BdG equations for bulk. (b) $\Delta_2$ as function of $\gamma$ for temperatures $T=2.64,\,2.65,\,2.7$ and $2.8\,{\rm K}$ from the BdG equations~(solid curves); the dashed curve shows the results of Eq.~(\ref{unif}) for $\Delta_2$ at $T \approx T_{c2}$; the dotted curve is the dependence given by Eq.~(\ref{unif_up}) at $T=2.7\,{\rm K}$. (c) The $\gamma$-susceptibility $\frac{\partial \Delta_2}{\partial\gamma}$ as a function of $T$ and $\gamma$, that diverges at the critical point $(T,\gamma)=(T_{c2},0)$.}
\label{fig1}
\end{figure*}

Equation~(\ref{exp_anom}) enables us to further examine the dependence of $\Delta_2$ on $\gamma$ for $T > T_{c2}$. When $\gamma \to 0$ and $T_{c2}< T < T_{c1}$ we find (for bulk, see \cite{suppmat})
\begin{eqnarray}
\Delta_2 = \gamma \Delta_{1,\gamma \to 0} \Big[n_2 \ln(T/T_{c2})\Big]^{-1},
\label{unif_up}
\end{eqnarray}
where $T_{c1}$ is the critical temperature of the stronger band for $\gamma=0$. The behavior of $\Delta_2$ changes dramatically as compared to Eq.~(\ref{unif}). The hidden critical point manifests here in the fact that the factor $\Big[n_2 \ln(T /T_{c2})\Big]^{-1}$ diverges when $T \to T_{c2}$ \cite{note1,suppmat}. We should note the existence of another (analogous) hidden criticality at $T_{c1}$, which is the reason why Eq.~(\ref{unif_up}) is not valid for $T=T_{c1}$. However, for $\gamma \to 0$ this point is in close vicinity to the overall critical temperature $T_c$ and, as a result, it is always strongly overshadowed by the usual critical behavior close to $T_c$.

{\itshape Numerical results.} In what follows, we perform a full numerical study within a microscopic formalism, in a broader range of temperatures and interband couplings. In addition, we need to confirm the above analytical results since Eq.~(\ref{exp_anom}) assumes the validity of the gradient expansion - which is only correct when the corresponding healing length goes to infinity when $\gamma
\to 0$. In other words, the important limitation of our $\gamma$GL-analysis is that we first assumed $\xi_2 \to \infty$ ($\gamma \to 0$) in order to subsequently obtain the diverging behavior of $\xi_2$ with the critical exponent $\mu$, which may be misleading.

As an appropriate theoretical formalism, we choose the Bogoliubov-de Gennes (BdG) equations~\cite{degen}. We emphasize that the standard two-band GL formalism cannot be used for our purpose, being well justified only near $T_c$. In the present case, the BdG equations read
\begin{eqnarray}
\left(
\begin{array}{cc}
T_i({\bf x}) & \Delta_i({\bf x})\\
\Delta^{\ast}_i({\bf x})
& -T^{\ast}_i({\bf x})
\end{array}
\right)
\left(
\begin{array}{c}
u_{i\nu}({\bf x})\\
v_{i\nu}({\bf x})
\end{array}
\right) = E_{i\nu} \left(
\begin{array}{c}
u_{i\nu}({\bf x})\\
v_{i\nu}({\bf x})
\end{array}
\right),
\label{BdG_2band}
\end{eqnarray}
where $u_{i\nu}({\bf x})$, $v_{i\nu}({\bf x})$ and $E_{i\nu}$ are the particle-like and hole-like wave-functions and the quasiparticle energy, with the subscript $i$ enumerating the band and $\nu$ the set of relevant quantum numbers. The single-electron energy reads $T_i({\bf x})=-\frac{\hbar^2}{2m_i}{\bf D}^2-\mu_i$, with $m_i$ the band mass (set to the free electron mass) and $\mu_i$ the chemical
potential measured from the lower edge of the corresponding band. Formally, the BdG equations for different band order parameters look decoupled but, in fact, they are connected through Eq.~(\ref{gap}) taken with
\begin{eqnarray}
\langle {\hat\psi}_{i\uparrow}({\bf x}){\hat
\psi}_{i\downarrow}({\bf x})\rangle=\sum\limits_{\nu} u_{i\nu}({\bf
x})v^{\ast}_{i\nu}({\bf x})\big[1-2f(E_{i\nu})\big], \label{anom}
\end{eqnarray}
where $f(E_{i\nu})$ is the Fermi distribution of quasiparticles. Inserting Eq.~(\ref{anom}) into Eq.~(\ref{gap}), one needs to remedy the ultraviolet divergence: though the product $u_{i\nu}({\bf x})v^{\ast}_{i\nu}({\bf x})$ vanishes for large energies, its decay is not sufficiently fast to provide convergence of the sum over the relevant quantum numbers given by Eq.~(\ref{anom}). Following the standard cut-off procedure as implemented in other papers on two-band superconductors~\cite{suhl,zhit,kogan}), the cut-off energy $\hbar\omega_c$ is used.

We first investigate a numerical solution of Eqs.~(\ref{BdG_2band}) and (\ref{anom}) for a bulk superconductor for small $\gamma$. Figure~\ref{fig1} shows calculated results for the weaker band in a bulk superconductor, for the set of parameters given in the figure. Note that this particular choice of the parameters is not decisive for our conclusions - similar results are found for $\mu_i$ up to $\sim 10\,{\rm eV}$, typical of metals. Our choice of $\mu_i \sim 10\,{\rm meV}$ is justified by recent angle-resolved photoemission experiments (ARPES) on ${\rm Fe}{\rm Se}_x{\rm Te}_{1-x}$, which revealed the presence of multiple bands with extremely small Fermi energies ($\sim 5\,{\rm meV}$) \cite{lub}.

Figure~\ref{fig1}(a) shows that the increase in $\Delta_2$ with $\gamma$ becomes faster as $T$ approaches $T_{c2} \approx 2.645\,{\rm K}$, which supports our expectations. $\Delta_2$ versus $\gamma$ in the uniform system is plotted in Fig.~\ref{fig1}(b) for selected temperatures, where a comparison is made with our analytical results, Eqs.~(\ref{unif}) and (\ref{unif_up}). As seen, for $T > T_{c2}$ and $\gamma \to 0$ the weaker-band order parameter $\Delta_2$ is indeed a linear function of $\gamma$. However, when $T$ approaches $T_{c2}$, this linear dependence is replaced by $\propto \gamma^{1/3}$, as confirmed by the very good agreement between the numerically obtained curves for $T=2.64\,{\rm K}$ and $2.65\,{\rm K}$ and the analytic curve for $T\approx T_{c2}$ in Fig.~\ref{fig1}(b). At $T=T_{c2}$ the critical behavior of $\Delta_2$ as function of $\gamma$ leads to a diverging $\gamma$-susceptibility, as shown numerically in Fig.~\ref{fig1}(c). Due to the well-known interrelation between the susceptibility and the coherence length, one expects that $\xi_2$ is also a divergent function of $\gamma$ at $T=T_{c2}$. However, to study this feature and check the scaling $\xi_2 \propto \gamma^{-1/3}$ found in the $\gamma$GL model, we need to abandon the uniform case and study a spatially varying two-band condensate.

\begin{figure}
\includegraphics[width=\linewidth]{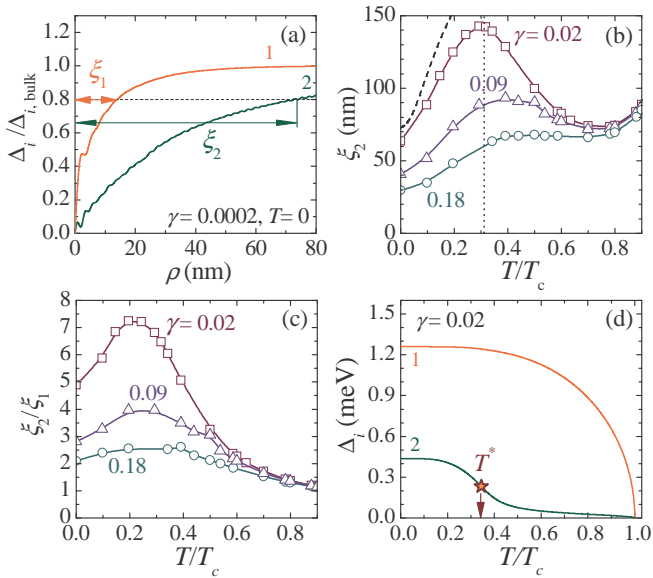}
\caption{(Color online) (a) $\Delta_i(\rho)$ in the vortex core as calculated for $T=0$ and $\gamma = 0.0002$ and the definition of the healing lengths $\xi_i$. (b) The numerically extracted healing length of the weaker band $\xi_2$ as a function of $T/T_c$ for different $\gamma$. The dashed curve represents $\xi_2$ in the independent-band limit $\gamma \to 0$, and the dotted line marks $T_{c2} =2.645\,{\rm K}$~(for comparison, they are shown versus $T/T_c$ with $T_c$ calculated for $\gamma=0.02$). (c) The ratio of the healing lengths $\xi_2/\xi_1$ versus temperature. (d) The bandgaps versus temperature and the definition of the crossover temperature $T^*$ (where $\xi_2$ peaks) extracted from the condition $\frac{\partial^2 \Delta_2}{\partial T^2} = 0$.\label{fig2}}
\end{figure}

We chose to investigate the single-vortex solution in a superconducting cylinder (with radius $R=300\,{\rm nm}$) by numerically solving Eq.~(\ref{BdG_2band}), and extract the band-dependent healing lengths $\xi_i$ from the size of the vortex core. When numerically studying a single-vortex solution in the cylinder, we follow a procedure similar to that of Ref.~\onlinecite{gygi}. We neglect the screening of the magnetic field, which is justified for extreme type-II superconductors. The spatial variation of the order parameters in two bands $\Delta_i(\rho)$ are shown  around a vortex in Fig.~\ref{fig2}(a) in units of their bulk  zero-temperature values $\Delta_{i,{\rm bulk}}$, for $\gamma = 0.0002$ and $T=0$. We defined the healing lengths by $\Delta_i(\rho=\xi_i)= 0.8 \Delta_{i,{\rm bulk}}$, i.e., $\xi_1=14\,{\rm nm}$ and $\xi_2=74\,{\rm nm}$ for this particular case~\cite{note2}. Figure~\ref{fig2}(b) shows the temperature dependence of $\xi_2$ for different $\gamma$ as extracted from our numerical results. We point out our main finding: the existence of a clear peak in $\xi_2$ at temperatures close to $T_{c2}$. Its peak value increases with decreasing $\gamma$ and almost ideally follows the scaling $\gamma^{-1/3}$ found within the $\gamma$GL model. For sufficiently small interband couplings, $\xi_2$ approaches its independent-band limit shown by the dashed curve in Fig.~\ref{fig2}(b). The peak in $\xi_2(T)$ also marks the temperature range where the most pronounced difference is found between the two healing lengths in the two bands [shown in Fig.~\ref{fig2}(c) for $\gamma < 0.2$]. When approaching $T_c$, the difference between the spatial profiles of the band-dependent condensates disappears, i.e., $\xi_2/\xi_1 \to 1$, in agreement
with Refs.~\cite{shan,kogan}. The present study shows that the effect of the
hidden criticality on the vortex core is very pronounced for $T_{c2} < 0.8T_c$. At higher temperatures the peak in $\xi_2$ is overshadowed by the usual critical behavior around $T_c$, unless the coupling $\gamma$ is extremely weak.

As seen in Figs.~\ref{fig2}(b,c), with increasing coupling $\gamma$ the temperature where $\xi_2$ peaks, labeled $T^*$, increases while the peak itself decreases (for the chosen parameters, the peak is washed out for $\gamma > 0.2$). Regarding the maximal peak magnitude, it is limited by the taken sample radius ($0.3\,{\rm \mu m}$). The largest value of $\xi_2$ in Fig.~\ref{fig2}(b) is well below this limit and finite size effects do not alter our conclusions. It is
interesting that $T^*$ can be evaluated without a time-consuming numerical study of the vortex solution, being very close to the temperature at which $\partial^2\Delta_2/\partial T^2=0$ for a bulk superconductor. This is illustrated in Fig.~\ref{fig2}(d) for $\gamma=0.02$, where the inflection point of $\Delta_2(T)$  at $T=$ 0.33 $T_c$ matches well with $T^*=$ 0.31 $T_c$ found from the peak position in panel (b). In either case, $T^*$ should be seen as the crossover temperature between the two regimes: lower temperatures - where $\Delta_2$ is governed by the properties of the weaker band (for sufficiently small $\gamma$), and higher temperatures - where $\Delta_2$ is controlled by the tunneling from the stronger band.

\begin{table}
\scalebox{0.82}{
\begin{tabular}{| c | c | c | c | c | c | c | c | c | c |}
\hline
Material & $\lambda_{11}$ & $\lambda_{22}$ & $\lambda_{12}$
& $T_{c1}(K)$ & $T_{c2}(K)$ & $T^*(K)$ & $T_c(K)$ & $\gamma$ & Ref.\\
\hline \hline
MgB$_2$ & 1.88 & 0.5 & 0.21  & 38 & 3.9 & 14 & 39 & 0.22 &
\cite{MgB2}\\
OsB$_2$ & 0.39 & 0.29 & 0.0084  & 2.1 & 1.2 & 1.5 & 2.1 & 0.074 &
\cite{OsB2}\\
LiFeAs & 0.63 & 0.642 & 0.061  & 17.7 & 6.7 & (14)  & 18 & 0.15 &
\cite{LiFeAs} \\
V$_3$Si & 0.566 & 0.472 & 0.0074  & 16.4 & 8.1 & 9 & 16.5 & 0.03 &
\cite{V3Si} \\
FeSe$_{1-x}$ & 0.482 & 0.39 & 0.001  & 8.3 & 3.1 &  3.2  & 8.3 & 0.005 &
\cite{FeSe} \\
\hline
\end{tabular}}
\caption{The coupling constants, calculated nominal critical temperatures $T_{c1}$ and $T_{c2}$ of the separate bands, the approximate crossover temperature $T^*$, the critical temperature $T_c$ and the interband coupling $\gamma$ for selected two-gap materials. The coupling constants are taken from the references in the last column of the table.
\label{table1}}
\end{table}

Finally, we briefly discuss available two-gap materials from the point of view of the hidden criticality. Based on the coupling constants available from Refs.~\cite{OsB2,MgB2,LiFeAs,V3Si,FeSe}, we estimate the dimensionless coupling strength $\gamma$ for various two-gap superconductors in Table~\ref{table1}~\cite{noteTable}. According to our calculations the best candidates to observe the effects associated with the hidden criticality seem to be V$_3$Si and FeSe$_{1-x}$ due to their particularly low values of coupling $\gamma$ and low $T^*$ and $T_{c2}$ in comparison with the bulk $T_c$. This is supported by the experimentally observed anomalies close to $T^*$ in these two materials, see Refs.~\cite{V3Si,FeSe}.

In summary, we demonstrated that the properties of a two-band superconductor are affected by {\it two critical points}. In addition to the ordinary critical temperature $T_c$, there exists a hidden critical point - at the critical temperature of the weaker band in the absence of coupling. Interband coupling $\gamma$ controls the proximity to this hidden critical point and governs the
criticality similarly to an external magnetic field for ferromagnetic materials or an external electric field for ferroelectric systems. For weak coupling, the weaker-band healing length exhibits an atypical temperature dependence, with the {\it well pronounced peak} close to the hidden critical point. This gives rise to a large disparity of healing lengths of the two condensates in this temperature region. Such a competition of length scales in a single material can lead to significant new physics, and may be closely related to the recently observed exotic behavior of vortex matter~\cite{GL,mosh}. Direct measurement of the weaker-band coherence length can be realized experimentally, as was demonstrated for $\pi$ band in ${\rm MgB}_2$~\cite{pi-band}. We suggest that more experimental work
should be done on recent multiband materials close to their hidden critical point, where evidence for criticality can be found through unusual thermal properties \cite{pop}, resurgence of fluctuations deep in the superconducting state or anomalous cusps in superfluid density as a function of temperature~\cite{V3Si,FeSe}. Further experimental advancements are facilitated by direct similarities of two-band superconductors with superconducting bilayers, where the atypical temperature behavior of the coherence length can be observed as a function of an
alterable interlayer coupling. Some signatures of such behavior are
already visible in Ref.~\onlinecite{bluhm}.

\begin{acknowledgments}
This work was supported by the Flemish Science Foundation (FWO-Vl). Useful discussions with A. V. Vagov are acknowledged.
\end{acknowledgments}

\section{Supplementary material}
\subsection{Derivation of the $\gamma$GL equation}

It is convenient to recast the self-consistency equation [i.e., Eq.~(1) of the article] for a two-band superconductor in the form
\begin{equation}
\left(
\begin{array}{c}
\Delta_1({\bf x})\\
\Delta_2({\bf x})
\end{array}
\right)=
\left(
\begin{array}{cc}
\lambda_{11}& \lambda_{12}\\
\lambda_{21}=\lambda_{12}& \lambda_{22}
\end{array}
\right)
\left(
\begin{array}{c}
R_1[\Delta_1({\bf x})]\\
R_2[\Delta_2({\bf x})]
\end{array}
\right),
\label{matr_self_A}
\end{equation}
where $R_i[\Delta_i({\bf x})]$ is a functional of $\Delta_i({\bf x})$ related to the anomalous Green function as [see Eq. (2) in the article]
\begin{equation}
R_i[\Delta_i({\bf x})] = \frac{\langle {\hat\psi}_{i\uparrow}({\bf x}){\hat
\psi}_{i\downarrow}({\bf x})\rangle}{N(0)}
\label{Ri}
\end{equation}
and $\lambda_{ij}=\lambda_{ji}=N(0)g_{ij}$, with $N(0)$ the total density of states. Then, Eq.~(\ref{matr_self_A}) can further be rearranged as
\begin{equation}
\left(
\begin{array}{c}
R_1[\Delta_1({\bf x})]\\
R_2[\Delta_2({\bf x})]
\end{array}
\right)=
\frac{1}{\eta}
\left(
\begin{array}{cc}
\lambda_{22}& -\lambda_{12}\\
-\lambda_{12}& \lambda_{11}
\end{array}
\right)
\left(
\begin{array}{c}
\Delta_1({\bf x})\\
\Delta_2({\bf x})
\end{array}
\right),
\label{matr_self_B}
\end{equation}
with $\eta=\lambda_{11}\lambda_{22}-\lambda^2_{12}$. Based on Eq.~(2) of the article and Eq.~(\ref{matr_self_B}) given above, one finds
\begin{equation}
\Big(\frac{\lambda_{11}}{\eta} -\alpha_2\Big)\Delta_2 + \beta_2\Delta_2|\Delta_2|^2 -{\cal K}_2{\bf D}^2\Delta_2 - \frac{\lambda_{12}}{\eta} \Delta_1= 0,
\label{gammaGL_A}
\end{equation}
where $\alpha_2$, $\beta_2$ and ${\cal K}_2$ are defined in Eq.~(3) of the article. Now, assuming that
$$
T=T_{c2}=\frac{2e^{\Gamma}}{\pi}\hbar\omega_c\,e^{-1/(n_2\lambda_{22})},
$$
with $n_i=N_i(0)/N(0)$ the partial density of states, and keeping only the terms linear in $\lambda_{12}$, Eq.~(\ref{gammaGL_A}) yields
\begin{equation}
\beta_2\Delta_2|\Delta_2|^2 -{\cal K}_2{\bf D}^2\Delta_2 - \frac{\lambda_{12}}{\lambda_{11}\lambda_{22}} \Delta_{1,\lambda_{12}\to 0}= 0,
\label{gammaGL_B}
\end{equation}
with $\beta_2$ and ${\cal K}_2$ taken at $T = T_{c2}$, which is the $\gamma$GL equation [$\gamma=\lambda_{12}/(\lambda_{11}\lambda_{22})$] given by Eq.~(4) of the article.

\subsection{Derivation of Eq. (7) in the article}

Let us again consider Eq.~(\ref{gammaGL_A}) but now for $T_{c2} < T < T_{c1}$ and {\it in the homogeneous case}. In this temperature domain $\Delta_2 \to 0$ for $\lambda_{12} \to 0$ and $\Delta_{1,\lambda_{12}\to 0} \not= 0$. From Eq.~(\ref{gammaGL_A}) we have
\begin{equation}
\Big[\frac{\lambda_{11}}{\eta} -n_2\ln\Big(\frac{2e^{\Gamma}\hbar\omega_c}{\pi T} \Big)\Big]\Delta_2 + n_2\frac{7\zeta(3)}{8\pi^2T^2}\Delta_2|\Delta_2|^2 = \frac{\lambda_{12}}{\eta} \Delta_1.
\label{gammaGL_C}
\end{equation}
Note that the expansion given by Eq.~(2) in the article and therefore also Eq.~(\ref{gammaGL_A}) given above is not justified for {\it a spatially nonuniform solution} at $T_{c2} < T < T_{c1}$ because the coherence length $\xi_2$ associated with a spatial variation of the condensate in a weaker band does not diverge as $\lambda_{12}$ goes to zero. In other words, one cannot invoke the gradient expansion in the Gor'kov derivation of Eq.~(2) in the article. However, for homogeneous case Eq.~(2) of the article is simply an expansion in powers of the order parameter $\Delta_2$ and, so, is fully correct. Keeping only terms linear in $\lambda_{12}$, from Eq.~(\ref{gammaGL_C}) one finds
\begin{equation}
n_2\ln(T/T_{c2})\Delta_2=\frac{\lambda_{12}}{\lambda_{11}\lambda_{22}}
\Delta_{1,\lambda_{12}\to 0},
\label{gammaGL_D}
\end{equation}
which is Eq.~(7) in the article.

We remark that at first sight Eq.~(\ref{gammaGL_D}) prescribes that $\Delta_2$ is infinite when $T \to T_{c2}$ and $\lambda_{12}= {\rm const}$. This is not true because the validity of the expansion in powers of $\lambda_{12}$ given in Eq.~(\ref{gammaGL_D}) requires that $\lambda_{12} \lesssim \lambda^{({\rm lim})}_{12}(T)$, and $\lambda^{({\rm lim})}_{12}(T) \to 0$ for $T \to T_{c2}$. Hence, $\lambda_{12}$ can not be considered as constant when using Eq.~(\ref{gammaGL_D}) for $T \to T_{c2}$. In particular, during the derivation of Eq.~(\ref{gammaGL_D}) we first assumed that
\begin{align}
\frac{\lambda_{11}}{\eta}-n_2\ln\Big(\frac{2e^{\Gamma}\hbar\omega_c}{\pi T} \Big) = & \,n_2 \ln\big(T/T_{c2}\big)\notag \\
&+\frac{\lambda^2_{12}}{\lambda_{11}\lambda^2_{22}} + {\cal O}(\lambda^4_{12})
\label{lambda_exp}
\end{align}
and, then, we ignored contributions of the order $\lambda^2_{12}$ and higher. However, when the first and second terms in the right-hand-side of Eq.~(\ref{lambda_exp}) become of the same order of magnitude, then the contributions $\propto \lambda^2_{12}$ can not be neglected any more. This makes it possible to estimate $\lambda^{({\rm lim})}_{12}(T)$ as
\begin{equation}
\lambda^{({\rm lim})}_{12}(T) \sim \lambda_{22} \Big[\lambda_{11} n_2 \ln\big( T/T_{c2}\big)\Big]^{1/2}.
\label{lambda_lim}
\end{equation}
which proves that $\lambda^{({\rm lim})}_{12}(T) \to 0$ when $T \to T_{c2}$. Notice that Eq.~(\ref{gammaGL_D}) does not hold for $T \geq T_{c1}$, where $\Delta_{1,\lambda_{12}\to 0} =0$. Formally, it gives $\Delta_2 = 0$ which means that $\Delta_2$ is not linear in $\lambda_{12}$ for small interband couplings any longer. This is directly related to the fact that $T=T_{c1}$ is one more hidden critical point for the system of interest, however it is now associated with a stronger band. As pointed out in the article, the hidden criticality associated with $T_{c1}$ is overshadowed by the usual critical behavior around $T_c$.


\begin{thebibliography}{99}
\bibitem{note_source} When keeping the direct analogy to, e.g., ferromagnet materials or ferroelectrics, the product $\gamma \Delta_{1,\gamma \to 0}$ must be viewed as the source field. Then, as $\Delta_{1,\gamma \to 0}$ does not depend on $\gamma$~(we keep the intraband couplings constant), it is possible to treat $\gamma$ as a source field measured in units of $\Delta_{1,\gamma \to 0}$.
\bibitem{note1} Note that similar arguments do not hold for $T < T_{c2}$ because $\Delta_{2,\gamma \to 0}$ is nonzero in this case, and the expansion given by Eq.~(\ref{exp_anom}) is no longer valid. However, it is reasonable to expect that $\Delta_2 -\Delta_{2,\gamma \to 0}$ is linear in $\gamma$ for $\gamma \to 0$ with a slope that increases when approaching $T_{c2}$.
\bibitem{note2} Our choice of the threshold $\Delta_{i} = 0.8\Delta_{i,{\rm bulk}}$ for the definition of $\xi_i$ is not essential for our conclusions. For example, when choosing $\Delta_i(\rho =\xi_i) = 0.9\Delta_{i,{\rm bulk}}$, we find $\xi_1 = 22\,{\rm nm}$ and $\xi_2=116\,{\rm nm}$, i.e. $\xi_i$ is simply rescaled by a factor of $1.56$-$1.57$. Note that the Friedel-like oscillations~(see, e.g., Ref. \cite{mach}) of $\Delta_1$ and $\Delta_2$ inside the vortex core [see Fig.~\ref{fig2}(a)] may cause more disperse values of the rescaling factor when decreasing the threshold down to $0.5\Delta_{i,{\rm bulk}}$. However, such oscillations are rapidly washed out at higher temperatures, see, e.g., \cite{mach}.
\bibitem{noteTable} The crossover temperature $T^*$ for MgB$_2$, OsB$_2$ and LiFeAs was estimated in a simplified model with two-spherical Fermi surfaces, taking into account realistic values of ratios of Fermi velocities and densities of states but otherwise neglecting subtle properties of the band structure. $\Delta_2(T)$ for LiFeAs is a concave function of temperature at all temperatures therefore value of $T^*$ is not defined. However, its second derivative narrowly approaches zero at 14 K, a value quoted in the table. The $T^*$ for V$_3$Si and FeSe$_{1-x}$ is obtained from the weak gap profiles in Refs.~\cite{V3Si} resp. \cite{FeSe}.
\end{thebibliography}
\end{document}